\documentclass[iop]{emulateapj}
\usepackage{graphicx}

\slugcomment{To appear in the Astrophysical Journal}
\shorttitle{A Pole-Dominated Corona on AB Dor}
\shortauthors{J.J.~Drake et al.}
\begin{document}
\title{X-ray Evidence for a Pole-Dominated Corona on AB Dor}
\author{Jeremy J. Drake\altaffilmark{1}, Sun Mi Chung\altaffilmark{1,2},
Vinay L. Kashyap\altaffilmark{1} and David Garcia-Alvarez\altaffilmark{1,3,4}}
\affil{$^1$Smithsonian Astrophysical Observatory,
MS-3, \\ 60 Garden Street, \\ Cambridge, MA 02138}
\affil{$^2$ Department of Astronomy,
The Ohio State University, 
4055 McPherson Laboratory, 
140 West 18th Avenue, 
Columbus, OH 43210-1173}
\affil{$^3$ Instituto de Astrofsica de Canarias, E-38205 La Laguna, Tenerife, Spain}
\affil{$^4$ Dpto. de Astrofsica, Universidad de La Laguna, E-38206 La Laguna, Tenerife, Spain}
\email{jdrake@cfa.harvard.edu}
\email{chung@astronomy.ohio-state.edu}
\email{vkashyap@cfa.harvard.edu}
\email{david.garcia@gtc.iac.es}

\begin{abstract}

A fine analysis of spectral line widths and Doppler shifts employing
Fourier transform and cross-correlation techniques has been applied to
{\it Chandra} HETG spectra obtained in 1999 October of the rapidly rotating young star
AB~Doradus in order to investigate its coronal topology.  The observation lasted 52.3~ks, covering 1.2 rotation periods.
The X-ray light curve obtained from integrating the dispersed signal revealed a
moderate intensity flare midway through the exposure in which the count rate increased sharply by about 50\%\ and subsequently decayed over the next 10~ks.  We find no
significant Doppler shifts in the spectra or modulation of the light curve that could be attributed to rotation of
dominant coronal structures at this epoch.  Individual spectral line widths are
statistically consistent with thermal broadening and formally require
no rotational broadening, while the $1\sigma$ limit to rotational
broadening corresponds to a
compact corona restricted to latitudes $>57\deg$.
Fourier analysis
suggests a small amount of rotational broadening is present
consistent with a corona restricted largely to the poles, and 
excludes models with surface rotational broadening or greater.  These
results present direct spectroscopic evidence that the
dominant coronal activity on rapidly-rotating active stars is 
associated with the dark polar spots commonly seen in photospheric
Doppler images, and support models in which these spots are of mixed
magnetic polarity that forms closed loops. 

\end{abstract}


\section{Introduction}
\label{s:intro}

The Sun is the only star for which we can presently image coronal
X-ray emission directly.  In order to understand how coronae on other
stars might be structured we must resort to indirect means, such as
eclipse or rotational modulation, and spectroscopic density
diagnostics.  The advent of the ``high resolution''
($\lambda/\Delta\lambda\sim 1000$) capability of the X-ray
transmission grating spectrometers on the {\it Chandra} X-ray
Observatory opened up further possible means of probing coronal
structure.  One of these is through the Doppler shifts of spectral
lines in the X-ray emitting plasma that arise as a result of orbital
or rotational motion.

For late-type stars, the velocities involved are typically of order
10-100~km~s$^{-1}$---easily sufficient for optical studies of
rotationally-driven changes in spectral line profiles that are now
routinely ``inverted'' to make quite detailed Doppler images of
surface features of varying brightness \citep[see,
e.g.,][]{Strassmeier:02}, and Zeeman Doppler images of surface magnetic features \citep[e.g.][]{Donati.Landstreet:09}. However, for instruments with a resolving
power of $\sim 1000$, such velocities only amount to a fraction of the
instrumental width, and Doppler studies are challenging.
Nevertheless, several studies have now demonstrated the utility of {\it Chandra} Low- and High-Energy Transmission Grating
(LETG,HETG) spectra for probing both orbital velocities and rotationally modulated Doppler shifts in magnetically active stellar systems, both binaries \citep[e.g.][]{Brickhouse.etal:01,Ayres.etal:01,Chung.etal:04,Huenemoerder.etal:06,Ishibashi.etal:06,Hussain.etal:12} and single stars \citep[e.g.][]{Hussain.etal:05,Drake.etal:08}.

In addition to detecting Doppler shifts in the {\it
Chandra} HETG spectrum of Algol caused by the orbital motion of Algol
B, \citet{Chung.etal:04} also found spectral line widths to have excess line broadening amounting to
approximately 150~km~s$^{-1}$ above that expected from thermal motion
and surface rotation.  The excess broadening implied that a
significant component of the corona at temperatures less than $10^7$~K
has a scale height of order the stellar radius.  A similar line width analysis applied to the rapidly rotating giant FK~Com by \citet{Drake.etal:08}  revealed Doppler shifts consistent with localized emission on one hemisphere with a scale height of about a stellar radius.   

The young K-type dwarf AB Doradus, with  a projected rotational
velocity of 90-100~km~s$^{-1}$
\citep{Rucinski:85,Vilhu.etal:87,Donati.Collier_Cameron:97}, presents a plausible, though challenging, case for
Doppler inference of coronal structure based on {\it Chandra} spectra.   \citet{Hussain.etal:05} found evidence for rotational Doppler shifts in {\it Chandra} Low Energy Transmission Grating (LETG) spectra, while \citet{Hussain.etal:07} combined {\it Chandra} data with Zeeman Doppler imaging (ZDI) of the stellar surface to infer coronal structure and scale height.  \citet{Cohen.etal:10} have also employed ZDI maps in order to infer coronal structure based on detailed magnetohydrodynamic (MHD) corona and wind models.

Here, we bring further, potentially more sensitive, analysis techniques to bear on
the {\it Chandra} High Energy Transmission Grating (HETG) spectra of AB~Dor.
The analysis partly follows methods developed in the study of Algol
HETG spectra by \citet{Chung.etal:04}, to which the reader is referred
for a more detailed technical discussion.  We use cross-correlation
and Fourier techniques to 
to determine where the bulk of the observed X-ray emission originates in this system at this epoch. 
AB Dor is introduced in the light of existing work in \S\ref{s:abdor}, and 
the {\it Chandra} observations and analysis are described in \S\ref{s:obs}.  The 
results of the analysis are discussed in \S\ref{s:discuss} before drawing the main
conclusions from the study in \S\ref{s:conclude}.

\section{AB Dor and its Coronal Activity}
\label{s:abdor}

AB Doradus (HD 36705) is a
relatively bright (V=6.9), rapidly rotating K0 star at a distance of about 15~pc, 
with a period $P=0.514d$ \citep{Pakull:81}. 
Its mass is estimated to be $0.86 M_\odot$ and it is in orbit with a
distant companion of mass $0.09 M_\odot$ \citep{Guirado.etal:97,Guirado.etal:10}.
Based on the similarity of its space velocity with that of the
Pleiades moving group 
\citep{Innis.etal:85}, and short period and high
Li abundance \citep{Rucinski:85,Vilhu.etal:87}, AB Dor was commonly
accepted as a young star coeval with the stars of the Pleiades ($\sim
100$~Myr) and that is just evolving onto the main-sequence.  However, the exact age of AB~Dor has been a topic of continuing controversy, with estimates ranging from as young as $30$~Myr to as old as 125~Myr 
\citep[e.g.][]{Collier_Cameron.Foing:97,Zuckerman.etal:04,Luhman.etal:05, Janson.etal:07,Torres.etal:08,da_Silva.etal:09,Guirado.etal:11,Barenfeld.etal:13,McCarthy.Wilhelm:14,Wolter.etal:14}.  The preponderance of these studies favor an age of at least 100~Myr or so.

The radius of AB~Dor was recently measured to be $0.96 \pm 0.06 R_\odot$ based on near infrared interferometry \citep{Guirado.etal:11}.
Commensurate with youth, rotational
modulation of the photometric light curve of AB~Dor, as well as
extensive optical Doppler imaging studies, suggest that its surface is
covered with large, cool starspots or ``maculae'', switching between active longitudes on a 5.5~yr cycle and probably  modulated with a cyclic period of about 20 years \citep[e.g.\ ][and references
therein]{Rucinski:85,Vilhu.etal:93,Donati.etal:99,Jarvinen.etal:05,Budding.etal:09,Arzoumanian.etal:11}.  The detailed distribution of these spots has been studied in surface brightness maps obtained over many years.  They show a persistent polar cap, together with 
variable mid- to low-latitude spots \citep[e.g.,][]{Donati.Collier_Cameron:97,
Jeffers.etal:07}.  The related technique of ZDI has been extensively employed on AB~Dor to study its surface magnetic field distribution and evolution.  The magnetic field maps reveal strong, complex field
distributions.  \citet{Donati.etal:99} reported at least 12 different
radial field regions of opposite polarities located all around the star from observations obtained in 1996 December.   They found a degree of five or greater for a spherical harmonics expansion of the underlying large-scale
poloidal structure. There is also a strong and often unidirectional azimuthal field
encircling the boundary of the polar cap found in the spot maps. 

There are several existing photometric and spectroscopic observational
clues as to the structure of the corona of AB~Dor.  Rotational
modulation studies based on EUV-X-ray observations of coronal emission
can, in principle, render some limited spatial information.  Evidence
for such modulation has been only very marginal in GINGA and EXOSAT
observations \citep{Collier_Cameron.etal:88,Vilhu.etal:93}, and essentially
absent in EUVE and ASCA 
data \citep{White.etal:96}.  A lack of strong rotational modulation
implies either one or more of the following: emission is distributed
fairly uniformly in azimuth; emission is extended with a scale height
of order the stellar radius or larger; emission is concentrated at the
stellar poles.  Later extensive monitoring with ROSAT revealed modest
$\sim 5-13$\%\ modulation in pointed observations and 19\%\ modulation
in all-sky survey data \citep{Kuerster.etal:97}, indicating the
presence of some lower latitude structure non-uniform in stellar
longitude.  \citet{Maggio.etal:00} reported an observation of a compact
flare (height $\la 0.3R_\star$) which took a rotation period to decay
and showed no evidence of self-eclipse, indicating a high latitude
location.  At lower chromospheric and transition region temperatures
($\sim 10^4$-$10^5$~K), rotational modulation was seen in the fluxes
of FUV lines \citep{Schmitt.etal:97,Ake.etal:00}.  More recently, \citet{Lalitha.Schmitt:13} did not find evidence for persistent rotational modulation based on {\it XMM-Newton} observations obtained over the past decade, and attributed observed variability to stochastic processes such as flares.

Perhaps more interesting are spectroscopic constraints.   \citet{Garcia-Alvarez.etal:05} found the coronal spectrum essentially identical to that of the Hyades tidally-locked close K1~V+DA binary V471~Tau that has been spun up to a very similar rotation period.  Both spectra exhibit a fairly broad range of coronal temperatures up to $10^7$~K or so, and coronal abundances that deviate from the solar mixture in according to an ``inverse First Ionization Potential" effect pattern that is typical of active stars \citep[see, e.g.,][]{Brinkman.etal:01,Drake.etal:01,Drake:03}.
\citet{Sanz-Forcada.etal:03} deduced a coronal scale height of less than the stellar radius based on electron densities estimated from {\it XMM-Newton} spectra, similar to density and optical depth constraints on scale height for other active stars \citep{Sanz-Forcada.etal:03b,Ness.etal:04,Testa.etal:04,Testa.etal:04b,Testa.etal:07}.

\citet{Hussain.etal:05} found evidence for
rotational Doppler shifts in lines of O~VIII observed using the {\it
Chandra} Low Energy Transmission Grating Spectrograph (LETGS) in 2000 December,
suggesting the presence of low-latitude compact emitting regions.
Analysis of HETGS 
Fe~XVII~$\lambda 15.01$ and FUSE Fe~XVIII ~$\lambda 974$ line profiles
indicated a coronal scale height of $\la 0.5$~$R_\star$.   
\citet{Hussain.etal:07} combined the {\it Chandra} observations with ZDI reconstructions of the surface magnetic field and its extrapolation into the corona, and concluded that coronal structures are no more than 0.3--$0.4 R_\star$ in height.  \citet{Cohen.etal:10} used a ZDI surface magnetic map based on spectropolarimetric observations obtained in 2007 December to drive a global MHD model of the wind and corona of AB~Dor.  They found the global structure of the stellar corona at lower latitudes was dominated by strong azimuthal wrapping of the magnetic field due to the rapid rotation.
\citet{Lalitha.etal:13} obtained simultaneous radio, optical, and X-ray observations covering a multi-component flare and found flaring filling factors of 1--3\%\ of the stellar surface. 

At cooler temperatures, rotational
broadening has been detected in lines of both C~III ($\lambda 977$,
$\lambda 1175$) and O~VI ($\lambda 1032$, $\lambda 1037$) in ORFEUS
and FUSE spectra \citep{Schmitt.etal:97,Ake.etal:00}.
\citet{Schmitt.etal:97} note that this does not exceed the
photospheric $v \sin i$ value, suggesting that the observed C~III and
O~VI emission is produced close to the stellar surface.   Based on later FUSE spectra obtained in 2003 December, \citet{Dupree.etal:06} infer 
activity at mid-latitudes with a scale height of $\sim1.3$--$1.4 R_\star$.   Optical spectroscopy has also revealed the presence of prominences, initially through discrete absorption features in H$\alpha$ \citep{Robinson.Collier_Cameron:86,Collier_Cameron.Robinson:89}, and subsequently also in the light of Mg~II and Ca~II \citep{Collier_Cameron.etal:90}.
These prominences comprise cool (8-10,000~K) clouds of
neutral, or near neutral, hydrogen at, and also perhaps beyond, the Keplerian co-rotation
radius.  \citet{Collier_Cameron.etal:90} estimated that between 5 and 20 prominences were present in 1988 December, each with a mass of a few $10^{17}$~g.  Ultraviolet spectroscopy with the HST Goddard High Resolution Spectrograph revealed narrow chromospheric line cores superimposed on very broad line wings, the latter being consistent with a location commensurate of that of the optically-detected prominences \citep{Brandt.etal:01}.  
Analogous structures have also been detected in the K dwarf component of V471 Tau based on C~II and C~IV absorption in FUSE spectra by \citet{Walter:04}.  Young G and K dwarfs of the $\alpha$~Persei cluster also show evidence of the same behavior, indicating that such prominences are likely a common feature of rapidly-rotating stars \citep{Collier_Cameron.Woods:92}.

All these observations provide a rather complex, and sometimes inconsistent, picture of the corona of AB~Dor.  In the following sections we examine the only existing {\it Chandra} HETG spectrum of AB~Dor, and the highest resolution X-ray spectrum of the star obtained to date that presents lines bright enough for detailed study, for further clues as to its outer atmospheric structure.

\begin{figure*}[h!]
\epsscale{1}
\plotone{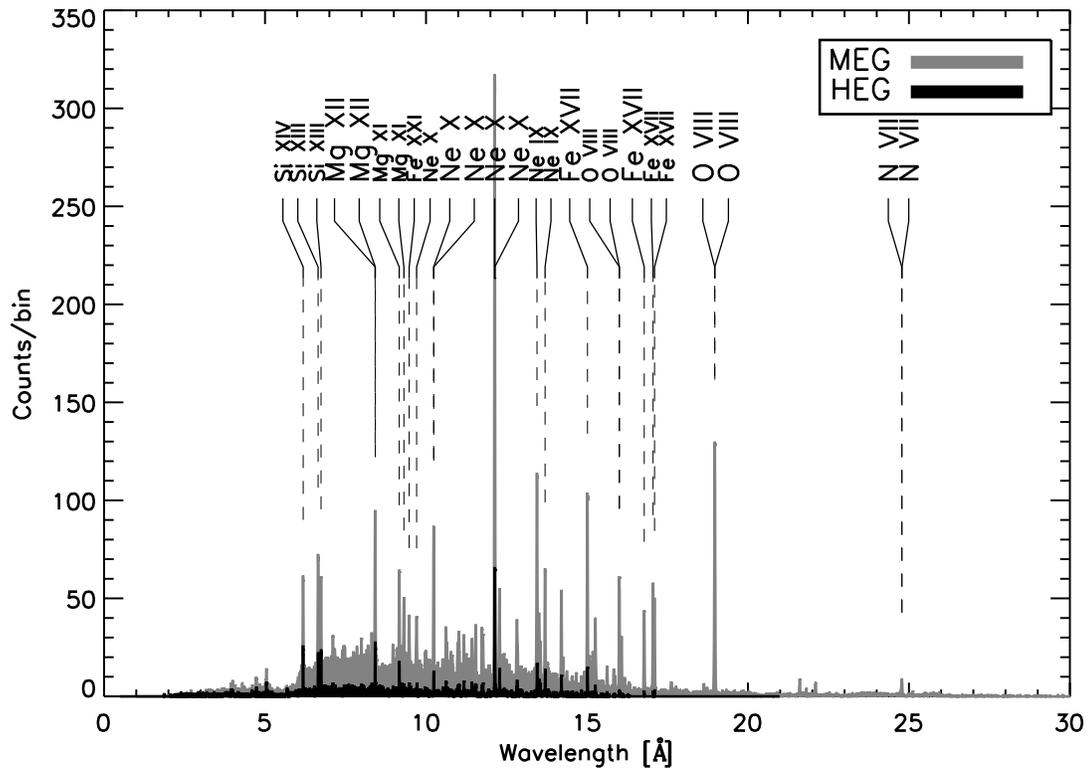}
\caption{The {\it Chandra} HETG spectrum of AB~Dor, for both HEG
  and MEG binned at 0.0025 and 0.005~\AA\ intervals, respectively.  
Emission lines that were used in our analyses are labeled
  in a larger font size than lines that were not used. }
\label{f:spec}
\end{figure*}

\section{Observations and Analysis}
\label{s:obs}

AB~Dor was observed using the {\it Chandra} High Energy Transmission
Grating (HETG) and ACIS-S detector on 1999 October 9 UT11:21 for a
total net exposure of 52.3ks, or approximately 1.2 stellar rotation
periods.  The data analyzed here were obtained from the Chandra public
archive\footnote{http://asc.harvard.edu/cda}.  The final
extracted spectrum showing the spectral lines used in this analysis is
illustrated in Figure~\ref{f:spec}.  Our analysis methods described
below, involving light curves, cross-correlation techniques and line
profile modeling, are similar to those described by
\citet{Chung.etal:04} and \citet{Drake.etal:08}, to which the reader seeking more detail is
referred.

\subsection{Light Curve}

Prior to performing detailed spectral analysis, the {\it dispersed}
photon events were first examined to ascertain the level at which
AB~Dor varied in X-rays during the observation.  Owing to the
ACIS-S 3.2s frame time, 0th order events are significantly affected by
photon pile-up and do not yield accurate photometry and are ignored
here.  Flares and rotational modulation caused by dominant
active regions are of particular interest since the location of their 
X-ray source might be betrayed by rotationally-induced Doppler shifts,
providing direct observational clues as to the locations of coronal
activity.  

\begin{figure}[h!]
\epsscale{1.2}
\plotone{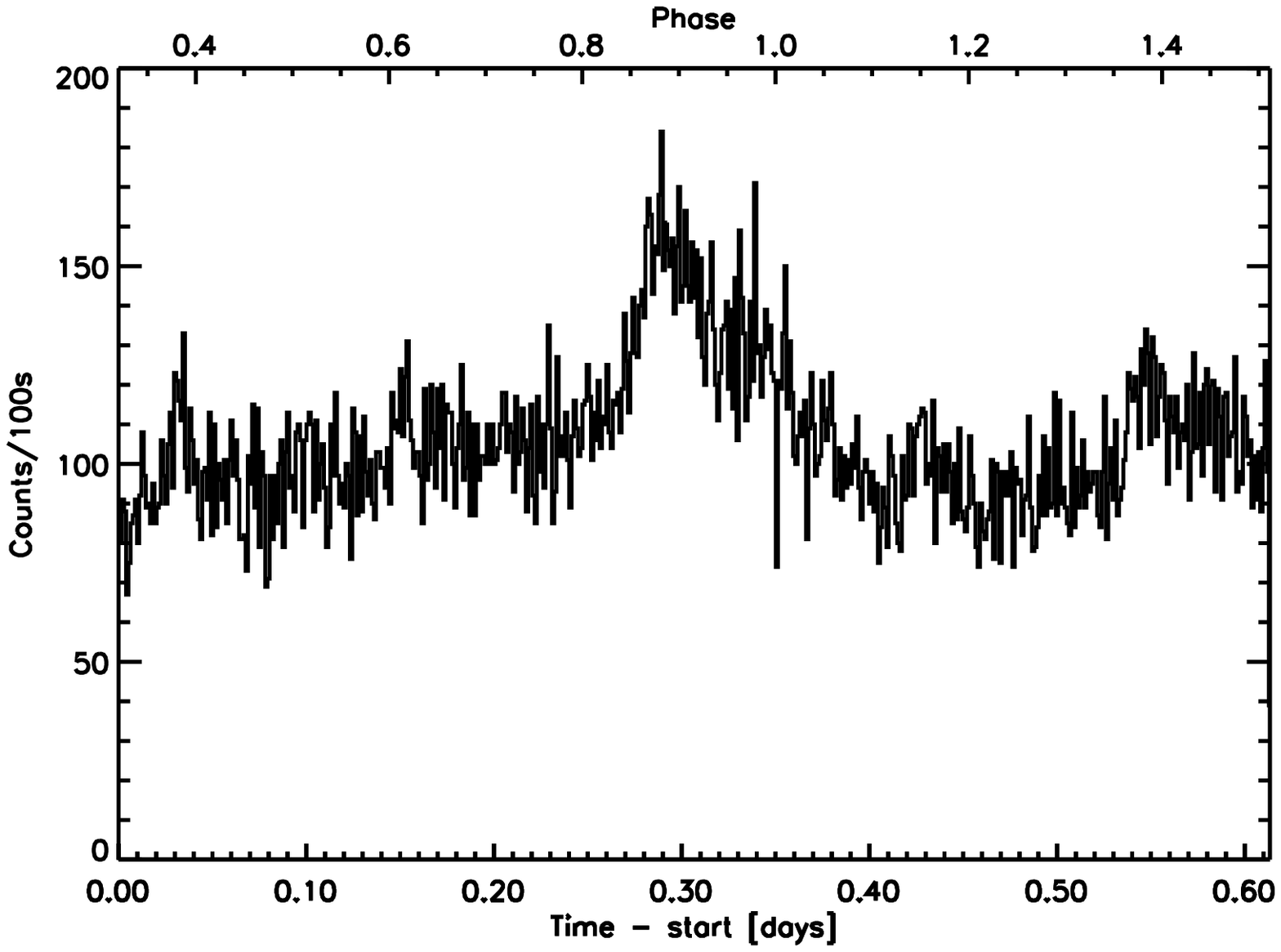}
\caption{The observed X-ray light curve of AB Dor derived from dispersed
  HEG and MEG events, plotted as a function of rotational phase.  The
  top axis shows the rotational phase computed with the ephemeris of
  \citet{Innis.etal:88}, and the bottom axis shows the elapsed time
  in days since the start of the observation.}
\label{f:lc}
\end{figure}

The X-ray light curve obtained from both HEG and MEG dispersed events
binned at 100s intervals is illustrated in Figure~\ref{f:lc}.  We also
show in this figure the rotational ``phase'' computed according to the
ephemeris of \citet{Innis.etal:88}, $HJD = 2444296.575 +0.51479E$.
There are several very small flare-like brightenings visible, but the
light curve is dominated by a moderate flare occurring midway through
the observation.  There is no obvious impulsive phase and the apparent
flare rise time is of order 2000s, which might be considered long for
a flare of this size whose decay timescale is of order 10000s.  It is
possible that an impulsive phase was occulted by the stellar disk just
prior to rotating into view, though Doppler shift measurements
discussed below suggest this might not be the case.  

While the light curve does not appear to show any structure that might be commensurate with rotational modulation,  the relatively short duration of the observation covering only slightly more than a single phase, combined with the flare, does hinder our ability to detect such modulation if it were present.   The flatness of light curve outside of the large and smaller flares indicates that the amplitude of any rotational modulation can only be of the order of 10\%\  or less.

\subsection{Phase-Related Doppler Shifts}

As noted above, Doppler shifts of spectral lines as a function of time
might betray the presence of particularly dominant coronal
structures.  The equatorial velocity of AB~Dor is about 95~km~s$^{-1}$,
which amounts to approximately
25\%\ of the width of the O~VIII 18.97~\AA\ Ly$_\alpha$ doublet as
seen by the MEG; such a shift should be easily detectable in the
brighter spectral lines of the AB~Dor soft X-ray spectrum.

\subsubsection{Analysis of Individual Spectral Lines}
\label{s:doppler}

In the first part of our analysis we searched for Doppler shifts of
individual spectral lines as a function of time.  Lines used in this
analysis are listed in Table~\ref{t:lines}.  These represent the
strongest lines in the HETG spectrum of AB~Dor longward of 8~\AA.
While H-like and He-like lines of Si are also reasonably bright, the
resolving power of the Chandra gratings is generally too low at these
wavelengths to be of significant additional value in this line-by-line
part of the study.

\begin{table}[h]
\begin{center}
\caption{A list of emission lines used in our analyses, including the
  rest wavelength, element and ion, and grating. }
\small
\begin{tabular} {c c c c}
\hline
Wvl [\AA] & Ion & Grating & Analysis$^\dagger$ \\
\hline
8.42  & Mg~XII & MEG,HEG & Doppler Shifts\\
12.13 & Ne~X & MEG,HEG & Both \\
14.20 & Fe~XVIII & MEG & Both \\
15.01 & Fe~XVII & MEG,HEG & Both \\
16.01 & O~VIII & MEG & Both \\
16.78 & Fe~XVII & MEG & Both \\
18.97 & O~VIII & MEG & Both \\
24.78 & N~VII & MEG & Both \\
\hline
\label{t:lines}
\end{tabular}
\vspace{-0.2in}
\end{center}
$\dagger$ Indication of whether the lines were used in Doppler shift or line
  broadening analyses.
\end{table}

Dispersed events within wavelength ranges of interest for our chosen
lines were binned over different time intervals, depending on the
number of counts in each line, so as to achieve approximately the same
signal-to-noise ratio in each bin.  Positive and negative orders were
co-added, then analyzed.   Line wavelengths were
measured by fitting modified Lorentzian functions (``$\beta$-profiles'') described by the
relation
\begin{equation}
F(\lambda)=a/[1+(\frac{\lambda-\lambda_0}{\Gamma})^2]^\beta
\label{e:lorentz}
\end{equation}
where $a$ is the amplitude and $\Gamma$ is a characteristic line
width.  For a value of the denominator exponent $\beta=2.5$, this
function has been found to be a good match to observed HETGS line
profiles, which are not well-matched by Gaussian profiles \citep{Drake:04}.
Fitting was performed using the {\it Package for INTerative Analysis of Line Emission (PINTofALE)}\footnote{PINTofALE is publicly available at  http://http://hea-www.harvard.edu/PINTofALE/ }  \citep{Kashyap.Drake:00} IDL\footnote{Interactive Data Language, Exelis Inc.} software routine {\sc fitlines}.   Uncertainties at the $1 \sigma$ level were determined from a thorough search of  the goodness-of-fit surface.

\begin{figure*}[h!]
\epsscale{1.2}
\plotone{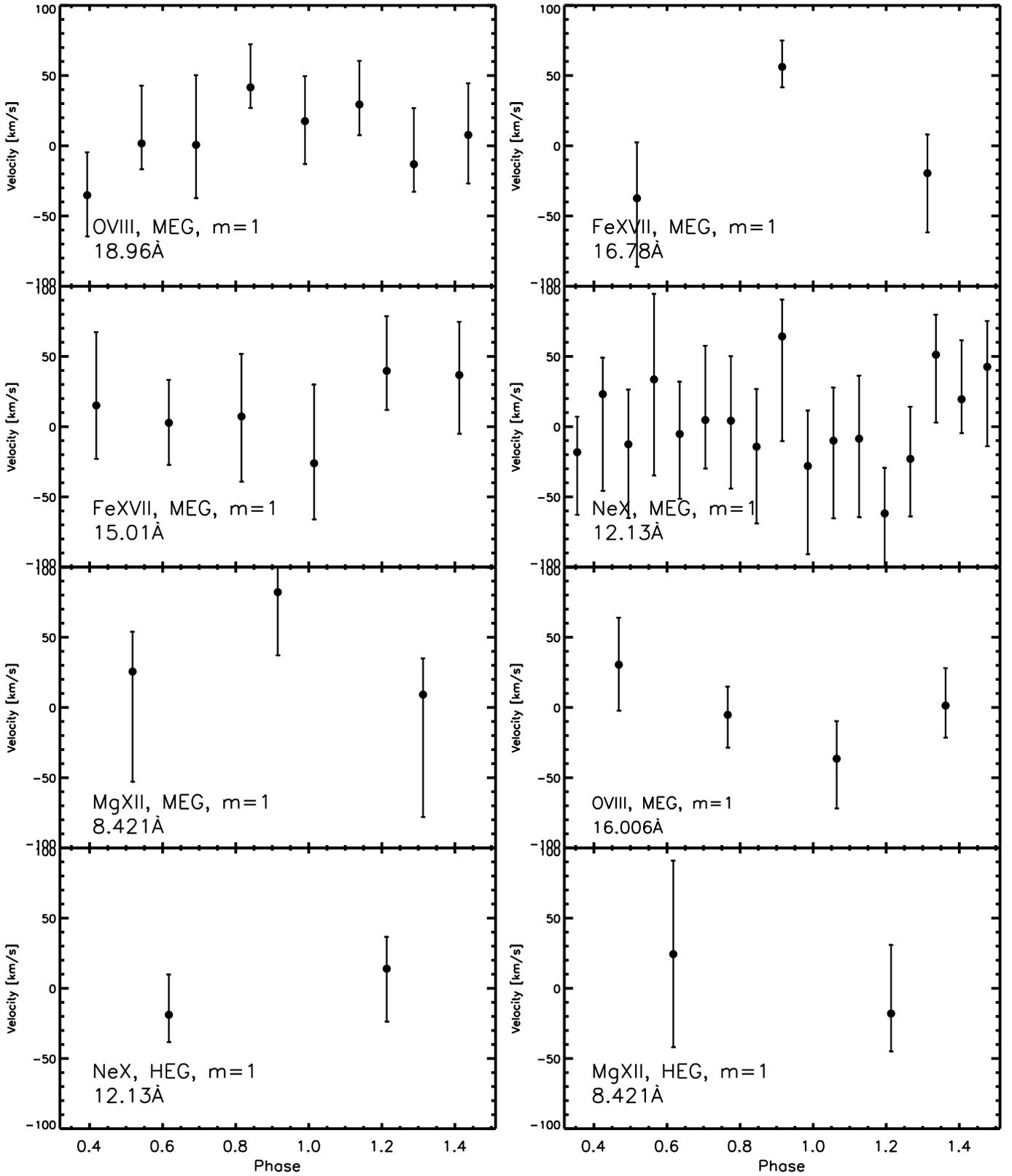}
\caption{The observed 
  line-of-sight orbital velocity and 1$\sigma$ error bars as a function
  of rotational phase for lines listed in Table~\ref{t:lines}, 
  expressed relative
  to the measured wavelength for the entire observation. }
\label{f:orbit_all}
\end{figure*}

Since the HETG observation covered slightly more than one full
rotation period, we determined line positions in the individual
time-filtered bins relative to the wavelengths measured for the same
lines from the entire observation.  In this way, we avoid
complications from small wavelength calibration errors or any
long-term drift in the dispersion relation.  The measured relative
line positions as a function of time are illustrated in
Figure~\ref{f:orbit_all}.  The resulting line positions are fairly
stable with little change as a function of orbital phase.  There is a
suggestion of a redshift at phase $\phi\sim 0.9$ in Fe~XVII~$\lambda 
16.78$ and Mg~XII~$\lambda 8.42$, but this signature is not seen in
other lines.  The O~VIII~$\lambda 18.96$ line also shows some evidence
for a secular trend with a single point near $\phi\sim 0.8$ exhibiting
a $2\sigma$ variation from the zero level, but this trend is opposite
that which one might be tempted to see in O~VIII~$\lambda 16.01$; the
latter is actually perfectly consistent with a zero net velocity
shift.

\subsubsection{Cross-correlation Analysis}
\label{s:crosscor}

In addition to measuring individual emission lines, we also utilized a
cross-correlation technique to obtain Doppler shifts as a function of
rotational phase.  The method involves comparison of the spectra
extracted for the different time bins with a reference spectrum in
order to determine whether or not there is a net Doppler shift between
the two.   This technique was first developed during 
earlier analyses of {\it Chandra} HETGS observations of Algol \citep{Chung.etal:04} and FK~Com \citep{Drake.etal:08}, and we refer the
reader to those works for a more detailed description.
The advantage of this type of cross-correlation technique
over the analysis of individual spectral lines is that it utilizes the
signal from the whole spectrum---strong and weaker lines, isolated and blended---and is therefore, at least in
principle, much more sensitive.  There are about 12 strong lines longward of 10~\AA\ in the MEG that have the highest velocity sensitivity and will provide the greater part of the signal, and a similar number in the HEG spectrum, although, again, all features in the spectrum contribute. 
The disadvantage of the technique is that it potentially confuses structures with different characteristic temperatures since it uses lines formed over a fairly large temperature interval.
We return to this in \S\ref{s:disc_results}.

Events were binned into five time intervals corresponding to different
rotational phases, and spectra were extracted for each phase bin.  These
spectra were then cross-correlated with the reference spectrum, which
in this case was the spectrum extracted from the entire exposure, in order to
determine the velocity shift at which our correlation signal
(represented by a $\chi^2$ statistic) was minimized.  As in the
analysis of Algol, the uncertainties in this procedure were estimated
using a Monte Carlo technique in which the analysis was repeated a
number of times for different Poisson realizations of the spectra.
The analysis was undertaken for both the HEG and MEG spectra separately.  The full range of each spectrum that contained significant signal was used.  These ranges were 2--26~\AA\ and 2--18~\AA\ for MEG and HEG, respectively.

\begin{figure}[h!]
\epsscale{1.2}
\plotone{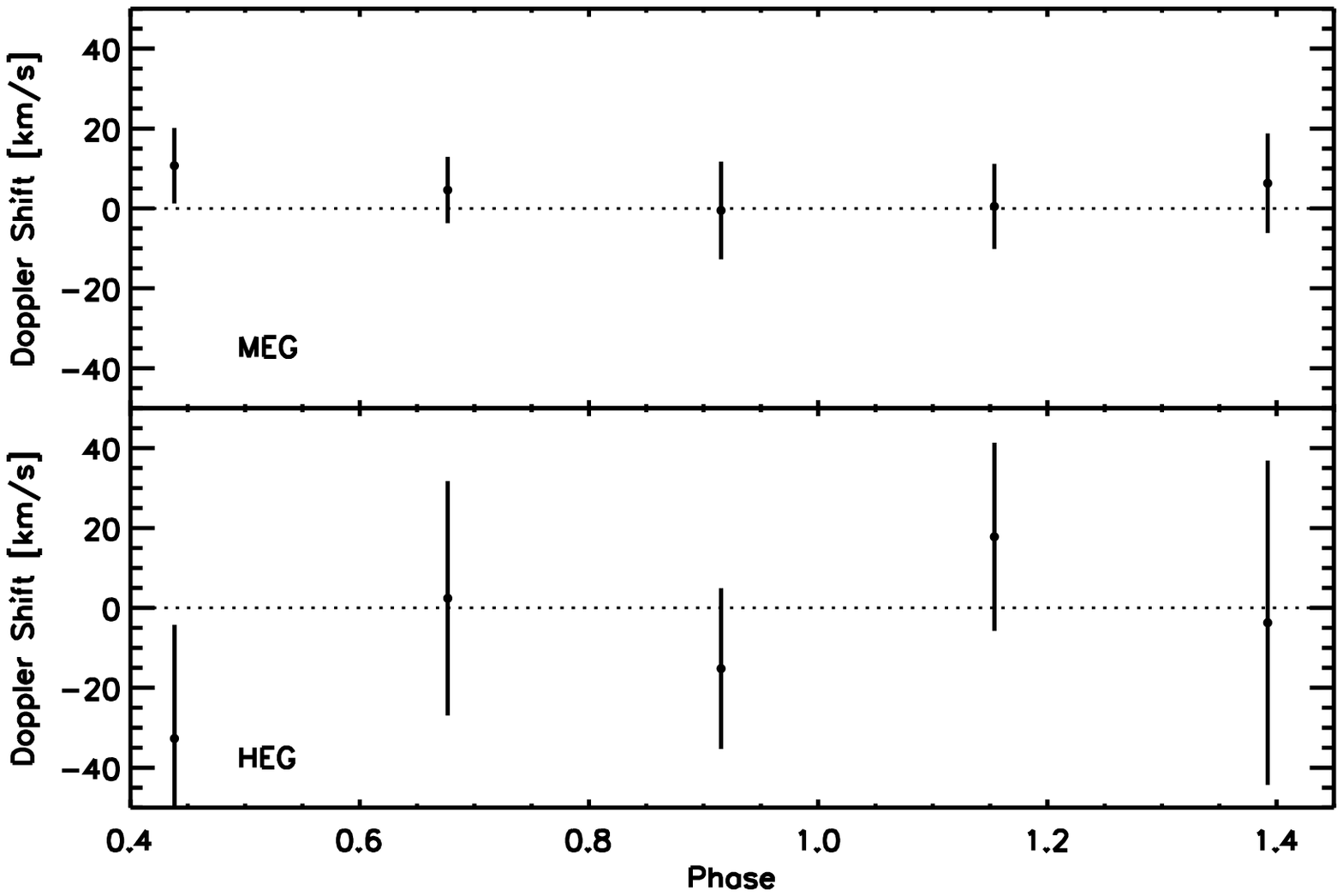}
\caption{The line of sight velocity as a function of
  rotational phase for MEG and HEG spectra, obtained by cross-correlating
  spectra from different time intervals with the spectrum from the
  entire observation.}
\label{f:crosscor}
\end{figure}

Figure~\ref{f:crosscor} illustrates the line-of-sight velocity of
AB~Dor as a function of rotational phase obtained from the
cross-correlation method, for the MEG (upper) and HEG (lower) data.
There are clearly no large velocity shifts detected, with the more sensitive, by virtue of the number of photon counts, MEG data placing quite stringent limits on average net velocity excursions of less than about 20~km~s$^{-1}$.  In the context of the projected surface equatorial rotational velocity of 95~km~s$^{-1}$, this lack of obvious systematic Doppler shifts 
indicates that
the corona of AB~Dor is not dominated by one or two low-latitude
active regions; the X-ray emission must be more evenly distributed
over the surface, or else resides predominantly toward polar regions
for which projected rotation velocities are small.

\subsection{Spectral Line Widths}
\label{s:widths}

The two scenarios mentioned above can in principle be diagnosed by
examining spectral line widths to search for excess rotation-related
broadening.  In our analysis, we compared line widths of observed data
to those of model profiles, which include the effects of thermal,
rotational, and instrumental broadening.  The procedure was again 
developed from that originally discussed in detail by \citet{Chung.etal:04} and \citet{Drake.etal:08}. 

In order to minimize uncertainties in the calibration of the
instrument line response function, we carried out the AB~Dor line
width analysis relative to an evolved star $\mu$~Vel, which is known
to have a very small projected rotation velocity of $v\sin
i=6$~km~s$^{-1}$ \citep{de_Medeiros.Mayor:95}.  Synthetic line
profiles were calculated for $\mu$~Vel based on the DEM obtained from
the {\it Chandra} HETG spectrum by \citet{Garcia-Alvarez.etal:06}.

Lines used in this analysis are listed in Table~\ref{t:lines}.  The
Mg~XII~$\lambda 8.42$ line was excluded in the line width
analysis because the dispersion at this wavelength is insufficient to
clearly discriminate between 
line widths of theoretical profiles with no
rotational broadening versus those broadened by surface rotation.  
Theoretical line profiles were
synthesized by convolving the thermal, rotational, and instrumental
line broadening components.  Details of theoretical line synthesis may
be found in \citet{Chung.etal:04}.  Thermal broadening was computed
using the differential emission measure as a function of temperature,
$\Phi(T)$, derived from the same HETG spectra by
\citet{Garcia-Alvarez.etal:05}.  For rotational broadening we assumed
solid body rotation of the stellar surface \citep[e.g.\ ][]{Gray:92},
corresponding to the projected rotational velocity $v\sin
i=95$~km~s$^{-1}$.  The stellar radius, $R_\star$, was taken to be
$1R_\odot$, which is consistent with the interferometric result $0.96\pm 0.06R_\odot$ 
of \citet{Guirado.etal:11}.  We also synthesized profiles for different effective
radii to simulate the effects of rotation of a corona with a scale
height significantly above the stellar surface.

\begin{table*}
\begin{center}
\caption{Observed and model line widths for AB~Dor and $\mu$~Vel. \label{t:widths}}
\begin{tabular}{ccccccc}
 ~~ & ~~ & \multicolumn{4}{c}{Full-Width at
Half-Maximum (\AA) } &
~~ \\ \cline{3-6}
 ~~ & Rest &  \multicolumn{2}{c}{Observation} & 
  \multicolumn{2}{c}{Model\tablenotemark{a} } & AB Dor excess
  \\ 
Grating & Wvl (\AA) & AB Dor & $\mu$ Vel
 & AB Dor & $\mu$ Vel & relative to $\mu$ Vel\tablenotemark{b} \\ \hline
MEG &12.13 & 0.021 $\pm$ 0.001 & 0.020 $\pm$ 0.001 & 0.023 & 0.020 & -0.001\\ 
MEG &14.20 & 0.021 $\pm$ 0.002 & 0.019 $\pm$ 0.002 & 0.023 & 0.020 & -0.001\\ 
MEG &15.01 & 0.021 $\pm$ 0.001 & 0.018 $\pm$ 0.001 & 0.022 & 0.019 & -0.000\\ 
MEG &16.01 & 0.022 $\pm$ 0.002 & 0.017 $\pm$ 0.003 & 0.024 & 0.021 &  0.002\\ 
MEG &16.78 & 0.022 $\pm$ 0.002 & 0.015 $\pm$ 0.001 & 0.023 & 0.020 &  0.005\\ 
MEG &18.97 & 0.024 $\pm$ 0.001 & 0.019 $\pm$ 0.003 & 0.026 & 0.024 &  0.002\\ 
MEG &24.78 & 0.027 $\pm$ 0.007 & 0.012 $\pm$ 0.006 & 0.029 & 0.026 &  0.011\\ 
HEG &12.13 & 0.014 $\pm$ 0.001 & 0.015 $\pm$ 0.002 & 0.015 & 0.013 & -0.003\\ 
HEG &15.01 & 0.011 $\pm$ 0.002 & 0.012 $\pm$ 0.003 & 0.013 & 0.011 & -0.003\\ 
\hline
\end{tabular}
\tablenotetext{a}{Model line widths include 
instrumental, thermal, and surface rotational broadening.}
\tablenotetext{b}{Excess line widths refer to 
$(w_{obs}-w_{mod})_{AB\; Dor}- (w_{obs}-w_{mod})_{\mu\; Vel}$, where
$w_{obs}$ and $w_{mod}$ refer to observed and model line widths,
respectively.}
\end{center}
\end{table*}

\begin{figure}[h!]
\epsscale{1.2}
\plotone{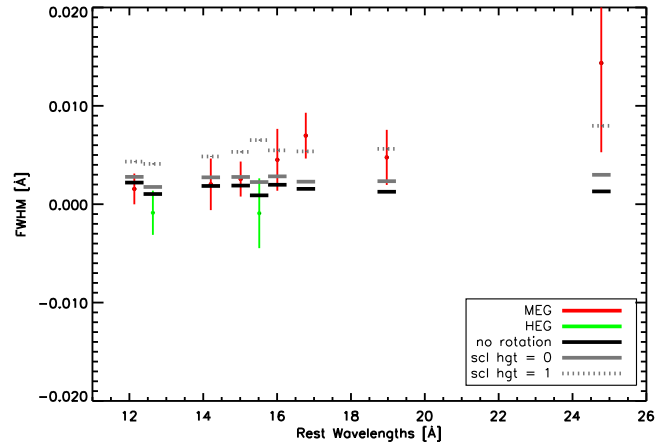}
\caption{Illustration of the observed line widths and $1\sigma$
  uncertainties of emission lines measured from the MEG and HEG
  spectrum of AB~Dor, with the corresponding $\mu$~Vel line widths
  subtracted.  
  The horizontal lines represent the model line widths of AB~Dor,
  relative to $\mu$~Vel.  
  The solid black horizontal line indicates predicted line
  widths based on a realistic optically-thin model, without rotational
  broadening.  The solid gray and dotted gray horizontal lines
  represent the predicted line widths for surface rotational broadening
  and rotational broadening at a scale height of $1R_\star$ above
  the stellar surface, respectively. }
\label{f:order_comp}
\end{figure}

Observed line widths were measured as described in \S\ref{s:doppler}.
Table~\ref{t:widths} lists the line widths of AB~Dor and $\mu$~Vel for
observed and model profiles, together with the 
AB~Dor excess widths
relative to $\mu$~Vel.  We also illustrate the AB~Dor excess line widths in 
Figure~\ref{f:order_comp}. The uncertainties for these excess line widths relative to $\mu$~Vel correspond to the quadrature propagation of the uncertainties of the individual AB~Dor and $\mu$~Vel line widths.
Also shown are the theoretical
line widths corresponding to the case of no rotation, rotation at the
stellar surface, and rotation at one stellar radius above the stellar
surface, all relative to $\mu$~Vel.  Here, we define the
scale height as the height above the stellar surface.  Note that,
under this definition, negative scale heights are possible and
correspond to the case where rotational broadening is less than that
induced by rotation of uniform surface emission.

\begin{figure}[h!]
\epsscale{1.2}
\plotone{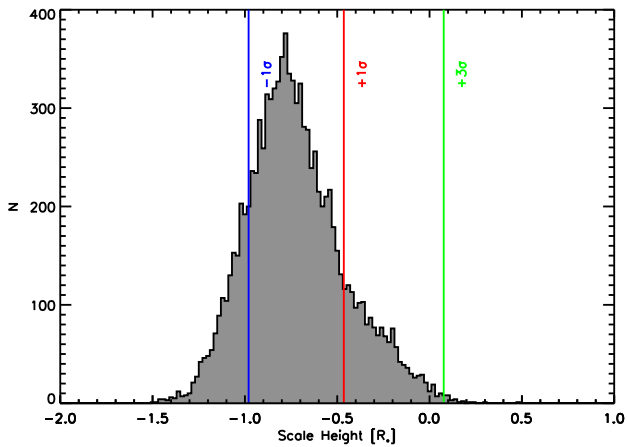}
\caption{The error weighted mean distribution of scale heights based
  on a Monte Carlo analysis of observed and predicted line widths of
  the MEG and HEG data, as described in text.  Scale heights are relative to the stellar surface and assume spherical symmetry, such that a height of zero corresponds to a uniform corona on the stellar surface with negligible vertical extension.  Negative scale heights imply the corona is restricted to high latitudes.
  The $1\sigma$ lower limit, along with the  $1\sigma$ and $3\sigma$
  upper limits are indicated by vertical lines.}
\label{f:height}
\end{figure}

In most cases the observed line widths in Figure~\ref{f:order_comp}
are consistent with the theoretical profiles corresponding to zero
rotation, within observational errors.  Most of the lines are also
consistent, within errors, with theoretical profiles that include
rotational broadening at the stellar surface.  
In order to determine to what extent the observations of individual
lines can constrain the scale height of the emission of AB~Dor, we
combined the individual results in the following way. For each AB~Dor
line width we subtracted the appropriate $\mu$~Vel
line width, then generated 100 Monte Carlo realizations 
within a normal distribution corresponding to the observed AB~Dor line
width (relative to $\mu$~Vel) and $1\sigma$ uncertainty. 
Line width realizations were interpolated onto a model curve that 
describes the line width as a function of scale height, to obtain a
distribution of scale heights for each line.  Finally, we computed the error
weighted mean of these distributions, which is shown in 
Figure~\ref{f:height}.  
The median scale height of this distribution occurs at -0.75$R_\star$,
with a $1\sigma$ lower limit of -0.98$R_\star$, and $1\sigma$ and
$3\sigma$ upper limits of -0.46$R_\star$ and 0.07$R_\star$,
respectively.  
Physically, this corresponds to the case in which emission is
concentrated toward the stellar poles.  The median scale height maps to a stellar surface latitude of $75\deg$ for very compact emission with respect to the stellar radius, while the $1 \sigma$ limit corresponds to $57\deg$ (the $3 \sigma$ upper limit allows very compact emission over the whole of the stellar surface).

\subsection{Fast Fourier Transform Analysis}

The general idea behind using Fast Fourier Transforms (FFTs) to probe
line broadening is to use the combined signal in the whole of the
spectrum, rather than just a small handful of bright, well-measured
lines.  Also crucial to the analysis of individual line widths is
accurate knowledge of the strength and location blends that might
significantly affect the observed line widths.  The analysis above was
based on single well-exposed lines for which blend contributions can
be estimated reasonably well, provided the temperature structure and
chemical composition of the corona are well-understood.  However, for
other lines somewhat weaker than our sample the blend contributions
become more significant and resulting systematic uncertainties in
measured line widths commensurately larger.

In a Fourier approach, it is the distribution of power that matters
and the exact placement of the myriad of weaker lines is less
important.  Similar Fourier techniques have been applied to high
resolution optical spectra of stars, as pioneered by D.~Gray and
co-workers \citep[e.g.\ ][]{Smith.Gray:76} are also in principle
capable of separating rotational broadening from other broadening
mechanisms such as turbulence with Gaussian velocity distributions.
However, {\it Chandra} grating spectra cover a decade in resolving
power from short to longer wavelengths and the broadening signatures
we seek amount to, at maximum, only a fraction of the instrumental
width and such fine differentiation will not be possible.  

Our method is to compare the power spectra of the observed AB~Dor HETG
spectra with those of synthetic spectra that have been broadened by
different rotation velocities.  Synthetic spectra were computed using
a similar approach to that described above for individual spectral
lines, using the emission measure distribution with temperature,
$\Phi(T)$, in addition to element abundances, of
\citet{Garcia-Alvarez.etal:05}.  Owing to the mass dependency of
thermal broadening, this needs to be accounted for separately for the
different elements prior to convolution with the rotational broadening
function and instrumental profile.

Since a given rotational broadening contributes more to an observed
line width at longer wavelengths where the instrumental resolving
power is highest, power spectra generated from synthetic spectra are
fairly sensitive to the relative intensities of the strongest lines
across the wavelength range.  While our simulated AB~Dor spectra
matched the observations very well in general \citep[see,
e.g.\ ][]{Garcia-Alvarez.etal:05}, some of the intensities of stronger Fe
lines, in particular, differed by $\sim 30$\%\ or so
from those observed.
These problems are largely attributable to difficulties and
complexities in theoretical predictions of the relative line strengths
themselves---a problem discussed at length in the case of Fe~XVII in
recent work of \citet{Doron.Behar:02} and \citet{Gu:03}.  We therefore
adjusted the simulated intensities of the strongest lines to match the
observed intensities.

Uncertainties resulting from Poisson counting statistics were
estimated for the FFT of the observed AB~Dor spectrum using a Monte
Carlo method.  We constructed 100 realizations of the observed
spectrum by randomizing the counts in each spectral bin within their
observed Poisson errors.  An FFT was then computed for each of the 100
realizations, and the $1~\sigma$ confidence limits on the mean FFT of
the observed counts spectrum were obtained from the resulting FFT
distributions.  In order to facilitate comparison between FFTs of
model and observed spectra, the FFTs were summed over intervals of 
10 bins, where the bin size was 0.005 \AA.

\begin{figure} [h!]
\epsscale{1.2}
\plotone{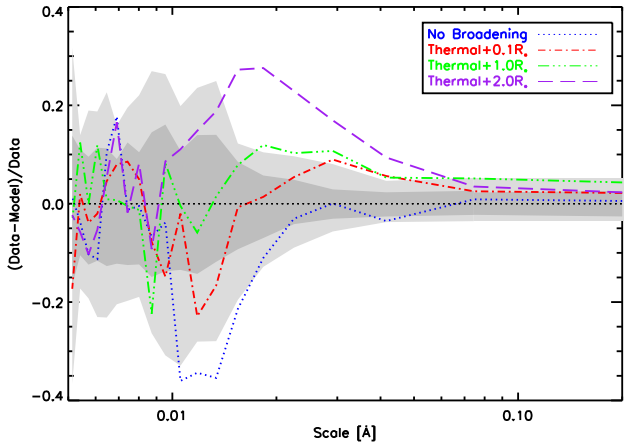}
\caption{Comparison between fast Fourier transforms of model and
observed spectra in the spectral scale size region of interest.  Model
spectra were computed as described in the text, and were convolved
with a rotational broadening function with different scale heights.
Here, the scale heights refer to the radius assumed in the rotational
broadening function, rather than to the additional height above
surface rotation as in the preceding figure and in the text (e.g.\
Thermal+$0.1R_\star$ implies a rotational broadening component
corresponding to a scale height of $-0.9R_\star$). The
shaded regions indicate the $\pm 1\sigma$ and $\pm 3 \sigma$
uncertainty bounds of the data obtained through Monte Carlo
simulations.}
\label{f:fft}
\end{figure}

The observed power spectrum and uncertainties are compared with those
of the synthetic spectra in Figure~\ref{f:fft}.  This analysis
indicates that spectra with only instrumental broadening are not a
good match to the observations, and some additional broadening is
required.  The FFT of the model spectrum with thermal broadening and
very small additional rotational broadening corresponding to a very
small scale height of $0.1R_\star$ (essentially equivalent to the case
of no rotational broadening) provides the best match to the
observations, although models with slightly larger scale heights---up
to $\leq 1R_\star$ or so cannot be excluded.  The model corresponding
to surface rotation scale height $1R_\star$ exceeds the $3\sigma$
error bounds at scales corresponding approximately to the line FWHM.
This indicates that the model line core is broader than observations
indicate.  Models with scale heights $> 1 R_\star$ are strongly
excluded.

\section{Discussion}
\label{s:discuss}

\subsection{Summary and Interpretation of Results}
\label{s:disc_results}

A careful analysis of {\it Chandra} HETG
spectra using
cross-correlation and FFT techniques has provided
quantitative constraints on the coronal scale height and structure of
the rapidly rotating corona on AB~Dor.  This information is encoded
in the observed spectrum through Doppler shifts of the source plasma.
These Doppler shifts amount to only a fraction of the instrumental
broadening and are not easily or reliably detected in single line
profiles or centroids.  However, the emphasis of the study presented here is
that the cross-correlation and FFT techniques we have exploited use the
signal in the whole of the observed spectrum and can be more
sensitive than analyses based on single lines.

Measured wavelengths of individual spectral lines illustrated in
Figure~\ref{f:orbit_all} show no obvious or significant Doppler shifts
relative to the mean frame of rest, although there is a hint of a
redshift at phase $\phi=0.9$ in lines of Mg~XII~$\lambda 8.42$ and
Fe~XVII~$\lambda 16.78$ that is also perhaps present in O~VIII~$\lambda
18.97$.  The phase
$\phi=0.9$ corresponds exactly to the peak of the flare evident in the
light curve in Figure~\ref{f:lc} and it is tempting to ascribe the weak velocity shift signature to it.  If so, it would imply the flare were situated on the receding hemisphere of AB~Dor. We noted in \S\ref{s:obs} that it
is possible that the impulsive phase of the flare was obscured by the
stellar disk.  Were this the case though we might have expected to see a
blueshift rather than redshift, as the flare approached along the
line-of-sight around the limb, and the lack of any net blueshift at
this phase tends to argue against such an explanation.  One might also
expect to see blue-shifted plasma as a result of chromospheric
evaporation within flaring loops on the visible hemisphere in which loops
are likely to be at least partly aligned along the line-of-sight (although to our knowledge no convincing blue-shifts due to chromospheric evaporation have been detected by {\it Chandra} to date).
That the redshift is not seen in the brighter
Fe~XVII~$\lambda 15.01$ line, however, suggests the redshift hint most prominent in  Fe~XVII~$\lambda 16.78$ is spurious with an origin in some degree of systematic uncertainty not accounted for in the analysis.

The cross-correlation of phase-resolved spectra provides a much more
sensitive measure of Doppler shifts than any of the individual lines
(Figure~\ref{f:crosscor}).  Figure~\ref{f:crosscor} shows that Doppler
shifts are almost exclusively within $1\sigma$ of 0~km~s$^{-1}$ and
are perfectly consistent with no significant net velocity shift.   
The lack of net Doppler shifts larger
than $\sim 20$~km~s$^{-1}$ implies that either coronal plasma
at low latitudes was relatively uniformly distributed in longitude, or 
that plasma was concentrated at the poles during the 1999 October epoch analyzed here.

We noted in \S\ref{s:crosscor} that the cross-correlation technique does not distinguish between lines formed at different characteristic temperatures.  In this sense, our measurements are essentially a signal-weighted average Doppler shift.  Since the longer wavelengths are also more sensitive to velocity,  the weighting for the MEG spectrum is going to be more toward the H-like Ne~X Ly$\alpha$ line and Fe~XVII and O~VIII features longward of this.  These ions are generally formed in the temperature range 2--$6\times 10^6$, although the increasing emission measure toward higher temperatures \citep[e.g.][]{Garcia-Alvarez.etal:05} means that weighting will be toward the high end of this range.  The HEG spectrum will be more sensitive to higher temperature features, including H- and He-like Si, Mg and Ne.  These are generally formed at temperatures of 5--$15\times 10^6$~K.    While the cross-correlation technique might mask some
subtle differences in radial velocity with temperature, the agreement of HEG and MEG results, and 
the fact that the individual
lines formed at different temperatures do not show any signs of
significantly different behavior, indicates that there is no temperature schism in the structure of the AB~Dor corona.

The degeneracy between a corona uniformly distributed in longitude or 
one concentrated at the poles indicated by the lack of significant Doppler shifts
can be removed by examining spectral line widths,
which we would expect to exhibit palpable broadening were the corona
to be uniformly distributed about the star, or concentrated at low
latitudes.  The line widths, however, show no evidence for strong
additional rotational broadening, in contrast to the HETG spectra of
Algol presented by \citet{Chung.etal:04}.  Both FFT and individual line width
analyses favor only very small additional broadening above that
induced by thermal velocities and the instrumental profile.  The
median coronal scale height from the Monte Carlo simulations
of all the lines combined (Figure~\ref{f:height}) is $h=-0.75R_\star$; ie
less than the stellar radius as measured from the stellar centre.
The $1\sigma$ upper limit is
$-0.46R_\star$---also less than the stellar radius.
Physically, these ``negative'' scale heights can be interpreted in
terms of the corona being restricted toward high latitude where
rotational broadening is smaller. Formally, the $1\sigma$ upper limit
corresponds to a corona limited to latitudes $> 57\deg$. 
Similarly, the FFT analyses also suggest the corona lies at restricted higher latitudes.  

These results provide direct spectroscopic evidence that the
dominant coronal emission in a rapidly-rotating active star such as
AB~Dor is largely associated with the dark polar spots commonly found in Doppler
imaging and ZDI analyses of photospheric spectra \citep[e.g.][]{Vogt.etal:99,Strassmeier:02,Jeffers.etal:05}.  
The results are not, however, definitive: the
$3\sigma$ upper limit to the scale height based on individual lines is
$h<0R_\star$ (surface rotation or less).   
A rigorous formal limit is more difficult to define
for the FFT analysis, but the data suggest strongly that $h<0R_\star$.
It is also not possible to rule out some small fraction of the emission as emanating from lower latitudes. Such emission would give rise to broadened line wings, of which we see no evidence in the spectra.  It is, however, possible that of the order of 10\%\ of the emission could be in a somewhat broader spectral component.

\subsection{Comparison with Other Epochs}
\label{s:disc_comparison}

The lack of phase-related Doppler shifts in the 1999 October spectra of AB~Dor contrasts with the 
modulation seen by \citet[][see also \citealt{Hussain.etal:07}]{Hussain.etal:05} in
the light of O~VIII~$\lambda 18.97$ in the LETG observation of 2002 December, which
exhibited a to peak-to-peak velocity change of 60 km~s$^{-1}$ and betrayed low-latitude 
structure during that epoch.  \citet{Hussain.etal:05} also investigated constraints on the coronal scale height from both {\it Chandra} LETG and FUSE spectra.  They found $h< 0.75R_\star$ (1$\sigma$) based
on the Fe~XVII~15.01~\AA\ line in the 2002 December epoch LETG spectra.  Based on FUSE observations from 1999 December and 2003 December of the
forbidden lines Fe~XVIII~$\lambda 974$ and Fe~XIX~$\lambda 1118$, they 
obtained $h=0.45\pm 0.3 R_\star$.  In principle, the
much higher FUSE resolution ($\lambda/\Delta\lambda\sim 20000$) should
render tighter constraints on line profiles than {\it Chandra}
spectra; unfortunately, however, the very weak forbidden Fe lines are
compromised by the presence of neighboring strong lines and line
blends.  

A salient point that the long FUSE observation, and the
{\it Chandra} LETG and HETG observations, were all obtained at significantly different epochs, should be borne in mind: while particular active longitudes apparently do exist on AB~Dor \citep{Jarvinen.etal:05}, details of the coronal topology
would be expected to change on shorter timescales and there is no
reason that particular features hinted at by one observation should be
present in another.  Our observations indicate that the bulk
of the coronal emission persists at high latitudes.  Both the
Doppler shifts and photometric rotational modulation found by 
\citet{Hussain.etal:05} suggest that some detectable emission can also
occur at lower latitudes.  This picture is supported by the pattern of
ambiguous EUV-X-ray rotational modulation evidence discussed in \S\ref{s:abdor}: 
while some observations found significant rotational
modulation, others, including the extensive study of \citet{Lalitha.Schmitt:13}, did not.

Further clues as to the temporal variation of the coronal emission of AB~Dor can be gleaned from Doppler and ZDI maps obtained in different years.  \citet{Donati.etal:03} presented both spot (Doppler imaging) and magnetic field (ZDI) maps for AB~Dor for epochs spanning 1998-2002.  All epochs exhibited an extensive dark polar spot covering latitudes above $70\deg$ or so \citep[see also][]{Jeffers.etal:07}, consistent with our inference of the location of the bulk of the coronal emission, and all showed some degree of structure at lower latitudes. 
However, qualitatively, there is quite a striking difference between maps presented for epoch 1999.97, which is fairly close to the epoch of our 1999 October observations, and those presented for 2001.99 in the sense that the latter contain much more extensive features and structure at mid-and low-latitudes.   Such structure is almost absent in the 1999.97 maps, even considering the more limited phase coverage of those observations.   This matches expectations from the 1999 and 2002 {\it Chandra} Doppler shift results reported here and by \citet{Hussain.etal:05}, respectively. 
The late 1990's also corresponds to the peak in brightness---and presumably the minimum in spot coverage---of AB~Dor based on the photometric monitoring of \citet{Jarvinen.etal:05}.

\subsection{Implications for Coronal Magnetic Field Extrapolation and Models}
\label{s:disc_implications}

As noted in \S\ref{s:intro}, the type of spectroscopic constraints on the
topography of coronal plasma reported here represent potentially important observational tests
of coronal structure extrapolated from ZDI maps of the photospheric magnetic field. 
Our deduction of predominately polar coronal activity on AB~Dor is in qualitative agreement with the 
ubiquitous presence of extensive polar spots on AB~Dor, and with such coronal field extrapolation 
by \citet{Jardine.etal:02b,Jardine.etal:02}, and \citet{Hussain.etal:02}. 
These authors inferred that, although the corona might be quite
extended, much of the model emission comes from high-latitude regions
close to the stellar surface.  

It has been especially noted by
\citet{McIvor.etal:03}, however, that ZDI cannot
reveal the nature of the magnetic field in polar regions and whether
field structures are predominantly closed or open.  These authors have
investigated the influence of different magnetic configurations on 
global coronal structure, while \citet{Mackay.etal:04} have performed magnetic flux transport simulations to study the formation of polar spots.  Both
Coriolis forces \citep{Schuessler.Solanki:92} and poleward meridional
flows \citep{Schrijver.Title:01} might contribute to the accumulation
of polar fields. The latter would tend to produce concentric rings of
alternating polarity, while the former might be expected to produce a mixed polarity pole.  \citet{Mackay.etal:04} find that both are actually needed to produce mixed polarity
regions. Images from ZDI support the idea of a mixed polarity
polar cap as the high latitude flux that  will ``feed" the cap invariably 
consists of intermingled positive and negative field.

In the models of \citet{McIvor.etal:03}, the Coriolis force and poleward meridional flow 
both produce coronae with dominant polar regions of closed field---regions
that would be bright in X-rays and consistent with our spectroscopic
constraints.  Alternative models constructed by adding
dipolar and unipolar field regions instead produce X-ray dark open field
regions at the poles and more equatorially-distributed coronae that are inconsistent with the spectroscopic evidence.  Such
models also imply rotational modulation in excess of observations
\citep{McIvor.etal:03,Jardine.etal:02b} and would also produce
rotationally-broadened line profiles well in excess of our observed
widths.

The coronal models reconstructed by \cite{Hussain.etal:07}  based on  ZDI contemporaneous with the LETG observations analyzed by \cite{Hussain.etal:05} had a significant high-latitude component, similar to what we find here, but also some structure at low latitudes, in qualitative agreement with the LETG spectra and X-ray light curve, and with what might be expected based on the surface imaging maps. 
\cite{Hussain.etal:07} noted that the potential field reconstructions of the coronal magnetic field depended on what was assumed for the surface magnetic field on the unseen pole (pointed away from the line-of-sight).  They discussed two approaches involving mapping the field on the visible pole to the invisible one, either reversing the sign of the field in the process or not.  The former produced much more rotational modulation of X-ray emission---and more than has been observed---than the latter, which was therefore preferred. 

\citet{Cohen.etal:10} have demonstrated that potential field extrapolation models for AB~Dor can be quite different to self-consistent models including the stellar wind and effects of rotation.  The wind acts to open up field lines that would otherwise remain closed without the additional wind forcing.  Rotational drag on the field was found to wind coronal structures quite tightly around the star, even below the Alfv\'en radius quite close to the stellar surface.  Another latent issue for coronal models of all types for active, rapid rotators based on magnetic maps was discussed in detail \citet{Arzoumanian.etal:11}, who considered the effects of dark spots on ZDI. 
Since this method is sensitive mainly to the magnetic field in the bright regions of the stellar surface, the magnetic signatures of the dark spots can be censored to some degree.  \citet{Arzoumanian.etal:11} modeled the magnetic field that might have been contained in these spots on AB~Dor and the similarly fast spinning, but fully-convective and lesser spotted, M4 dwarf V374~Peg by adding magnetic flux to the magnetograms.  The effect of missing flux for the latter star was fairly small, but very significant for AB~Dor that hosts a large, seemingly permanent, polar spot.  Adding polar flux of a single polarity significantly increased the model plasma emission measure as well as the rotational modulation, the latter well beyond observed limits.   The correlation between X-ray luminosity and large-scale magnetic flux assessed using ZDI presented by \citet{Vidotto.etal:14}, covering several orders of magnitude in both quantities, suggests that the loss of emission measure leading to X-ray emission is probably not catastrophic and perhaps also provides some support for mixed, rather than single, polarity polar fields.

The spectroscopic study presented here again tends to support a more mixed polarity field at high latitudes, which should diminish the effects of any clumpy, azimuthally non-uniform emission at lower latitudes.  The results could also be important for understanding angular momentum loss on young, rapid rotators.  The presence of missed polarity in the dark polar spots will act to close down field lines there that would otherwise be expected to present a strong source of wind-driven mass-loss.  Alternatively, the polar regions could be a copious source of flares and coronal mass ejections (CMEs) that might dominate the mass loss of very active stars \citep[e.g.][]{Drake.etal:13}.  The effect of the latter on angular momentum loss will be difficult to assess without greater understanding of such CME events, their opening angle,  and at what distance from the star they become super-Alv\'enic.

The \citet{Arzoumanian.etal:11} study presents a case for caution in interpretation of observed magnetograms for use in coronal models due to the inherent limitations of the observational technique.  In addition to dark spot field censoring, limited spatial resolution can also give rise to missing flux of mixed polarity on scales similar to, or smaller than, the resolution limit.  This missing flux and smearing of surface magnetic field structure can also give rise to significantly different predictions of coronal structure and emission \citep{Johnstone.etal:10,Garraffo.etal:13}.  
Spectroscopic studies such as that presented here, together with diagnostics of coronal plasma density and optical depth, are able to provide valuable observational leverage to overcome these difficulties, and potent tests of the properties of coronal models. 

\section{Conclusions}
\label{s:conclude}

We have demonstrated that {\it Chandra} HETG spectra are sufficiently
sensitive to provide Doppler shift constraints on the coronal
structure of AB~Dor.  Based on a fine analysis of individual spectral
lines, and on cross-correlation and Fourier techniques, we draw the
following conclusions.

\begin{enumerate}

\item In contrast to the finding of X-ray photometric rotational
modulation and rotation-driven Doppler shifts in the light of the
H-like O~VIII Ly$\alpha$ resonance line based on LETG observations
obtained in 2002 December by \citet{Hussain.etal:05}, we see no
significant rotational modulation or Doppler signatures in the 1999
October spectra analyzed here.  A cross-correlation analysis of both
HEG and MEG spectra places stringent limits on phase-related Doppler
shifts of $< 20$~km~s$^{-1}$.  These results are qualitatively consistent with surface images 
that appear to show less low-latitude structure in 1999, and with observations of variable degrees of low-latitude spot coverage at different epochs \citep{Jeffers.etal:07}.
There is no evidence in the spectra for
significant blueshifts associated with a modest flare that occurred
toward the middle of the observation, which might be expected from
chromospheric evaporation in loops aligned along the line-of-sight.

\item Some individual spectral lines exhibit Doppler shifts which
appear to be statistically significant, but which are not seen in
other spectral lines; uncertainties in observed line positions
therefore contain an additional element of uncertainty above that of
the formal statistics of profile fitting.  As also noted by
\cite{Chung.etal:04}, such additional
uncertainties are likely to be associated with the stability and ultimate precision of the
{\it Chandra} optical system.

\item Individual line widths together with Fourier analysis of the
whole spectrum are consistent with model profiles with very little
additional rotational broadening, indicating that the emission
originates at the stellar poles.  These results then provide the 
direct spectroscopic evidence that the dominant coronal activity on
rapidly-rotating active stars is associated with the dark
polar spots commonly seen in photospheric Doppler images.  The lack of
significant rotational broadening in coronal lines support models in
which these spots are of mixed magnetic polarity and able to sustain
closed coronal loops, rather than of dipolar or unipolar origin.

\end{enumerate}

The sensitive analysis techniques developed by \citet{Chung.etal:04} for Algol and furthered here for AB~Dor provide powerful means for
investigating coronal structure of nearby rapidly-rotating stars.

\acknowledgments

We thank an anonymous referee for a very helpful report that enabled us to significantly improve the manuscript. 
We thank the NASA AISRP for providing financial assistance for the
development of the PINTofALE package, and the CHIANTI project for
making publicly available the results of their substantial effort in
assembling atomic data useful for coronal plasma analysis.  
JJD and VLK were supported by NASA contract NAS8-03060 to the {\em Chandra
X-ray Center} during the course of this research.  DG was supported by
{\it Chandra} grants GO1-2021, GO2-3010X and AR4-5002X.  



\end{document}